\documentclass[preprint,showpacs,amsmath,amssymb]{revtex4}
\usepackage{graphicx}

\usepackage{bm}
\usepackage{xy}
\usepackage{epsfig}
\usepackage[usenames]{color}

\begin{document}
\def\1{\'{\i}}

\title{Multifractal analyses of row sum signals of elementary cellular automata
%A comparison study of three scaling methods for the (90,165) and (150,105) complementary pairs of ECA rules
}

\author{J.S. Murgu\'{\i}a$^{1,2}$, H.C. Rosu$^1$\footnote{Corresponding author.  Electronic mail: hcr@ipicyt.edu.mx \hfill CA-Comp2.tex}}
\affiliation{$^1$ IPICyT, Instituto Potosino de Investigacion Cientifica y Tecnologica, \\ Apdo. Postal 3-74 Tangamanga,
78231 San Luis Potosi, S.L.P., Mexico\\
$^2$
Universidad Aut\'onoma de San Luis Potos\1, \'Alvaro Obreg\'on 64, 78000 San Luis Potos\1, M\'exico}

%\author{H.C. Rosu\footnote{Corresponding author.  Electronic mail: hcr@ipicyt.edu.mx}}
%\affiliation{IPICyT, Instituto Potosino de Investigacion Cientifica y Tecnologica, \\ Apdo. Postal 3-74 Tangamanga,
%78231 San Luis Potosi, S.L.P., Mexico}

\begin{abstract}
\noindent We first apply the WT-MFDFA, MFDFA, and WTMM multifractal methods to binomial multifractal time series of three different binomial parameters and find that the WTMM method indicates an enhanced difference between the fractal components than the known theoretical result. Next, we make use of the same methods for the time series of the row sum signals of the two complementary ECA pairs of rules (90,165) and (150,105) for ten initial conditions going from a single 1 in the central position up to a set of ten 1's covering the ten central positions in the first row. Since the members of the pairs are actually similar from the statistical point of view, we can check which method is the most stable numerically by recording the differences provided by the methods between the two members of the pairs for various important quantities of the scaling analyses, such as the multifractal support, the most frequent H\"older exponent, and the Hurst exponent and considering as the better one the method that provides the minimum differences. According to this criterion, our results show that the MFDFA performs better than WT-MFDFA and WTMM in the case of the multifractal support, while for the other two scaling parameters the WT-MFDFA is the best. The employed set of initial conditions does not generate any specific trend in the values of the multifractal parameters.
%\\
%{\em Keywords:} \\
%\\
%%{\em 2010 Mathematics Subject Classification:} 81Q60
\end{abstract}

%\pacs{03.65.Ge, 03.65.Fd, 02.30.Hq}
\pacs{\\
{\tt 89.75.Da} -- Systems obeying scaling laws \\
{\tt 05.45.Tp} -- Time series analysis\\
{\tt 05.40.-a} -- Fluctuation phenomena, random processes, noise, and Brownian motion
}

\maketitle

\bigskip

\medskip

\bigskip

\section{Introduction}

Around 1975, Mandelbrot coined the term fractals to dynamical systems and objects with a hierarchy of structures that can be detected under changes of scales although related mathematical research on self-similar entities can be clearly traced back for at least two centuries and even some formulations in Euclid's Elements can be judged to allude to self-affine properties. If the structures that appear at lesser scales are identical to the bigger ones the systems or objects are monofractals, while if the structures are different on various scales they can form the more general class of multifractals. It is then natural to think of multifractals as complex systems but with a complexity that can be still characterized in terms of scaling power laws whose exponents are known as scaling parameters. In addition, during the last two decades of the past century the paradigms of self-organized criticality and $1/f$ noise helped to make firm the idea of scaling laws of many hierarchical processes. Moreover, such scale-dependent structures can be measured only with measuring devices that are scale dependent themselves and their complexity can be characterized in terms of non-integer dimensions such as the Hausdorff dimension.

%Time series containing possible nonstationary signals are a common way of recording the dynamics of complex systems characterized by scaling properties and their fractal dimensions \cite{hp}.
On the other hand, wavelet transforms (WT, briefly introduced in the Appendix) are pervading the analysis of multifractal processes since about two decades.
It is well known that the spectra of Hausdorff dimensions are accessible from time series by a Legendre transform of the corresponding Hentschel-Procaccia (HP) spectra \cite{hp}.
Muzy and collaborators \cite{muzy1} proposed a dimension type characteristic based on the wavelet transform of a measure and
in \cite{arne1} they demonstrated through several examples that the wavelet formalism yields the same spectra as the HP formalism and raised the conjecture that the equivalence of both formalisms should be valid in more general situations.

In this paper, we are interested in testing several multifractal numerical procedures to determine
the scaling properties of the time series generated by the evolution of elementary cellular automata (ECA), originally introduced by von Neumann and Ulam during the 1960's under the name of `cellular spaces' as a way of modelling biological self-reproduction. ECA are a class of discrete dynamical systems since they evolve in discrete time steps.
Willson \cite{will84} proved a long time ago that ECA are fractals by providing rigorous definitions and calculations of their fractal dimensions.
In 2003, Sanchez \cite{s03} was the first to study the multifractality of some ECA evolution rules using the random walk approach of Peng et al \cite{Pengetal} and together with Alonso-Sanz also examined the interplay between multifractality and memory effects \cite{sas}. In 2005, Nagler and Claussen \cite{NC} systematically computed the power spectra of the ECA time series of their row sum signals and found spectra of the type $1/f^{\alpha_p}$ with $\alpha_p=1.2$ and $\alpha_p=1.3$ but only for 10\% of the 256 ECA rules.

This paper is organized as follows. ECA are introduced in the next section. A brief description of the numerical scaling methods is given in the third section. The multifractal scaling analyses of the ECA time series for the so-called row sum signals makes the body of the fourth section and last come the conclusions.

\section{Cellular Automata (CA)}

During 1980s, because of the impetuous research of Wolfram cellular automata turned into a new scientific paradigm related to their ability to mimic through simple computed rules many natural processes \cite{W1,W2}.
The state space of a CA of size $N$ is the set $\Omega=\mathbb{Z}^N_k$ of all sequences of $N$ cells that take values from $\mathbb{Z}_k=\{0,1,\ldots,k-1\}$, where its evolution is defined by the repeated iteration of an evolution operator $\mathcal{A}:\mathbb{Z}^N_k \to \mathbb{Z}^N_k$.
 In this paper, we consider only two-digit cells, $\mathbb{Z}_2=\{0,1\}$, regularly distributed in one space and one time dimensions,
 in which case the CA are called elementary. These ECA have only $2^3=8$ different first-order neighbourhoods $x_{i-1}, x_{i}, x_{i+1}$ which generates $2^{2^3}=256$ possible rules that are counted from 0 to 255.
 An automaton state $\underline{x} \in \mathbb{Z}^\mathbb{Z}_2$ has coordinates
 $(\underline{x})_i = x_i \in \mathbb{Z}_2$ with $i\in \mathbb{Z}$, and the automaton state at time $t\geq 0$ is denoted
 by $\underline{x}^t \in \mathbb{Z}^\mathbb{Z}_2$ and its evolution is defined iteratively by the rule $\underline{x}^{t+1} = \mathcal{A}(\underline{x}^t)$. Starting from the initial state $\underline{x}^0$, the automaton
 generates the forward space-time pattern $\mathbf{x} \in \mathbb{Z}^{\mathbb{Z} \times \mathbb{N}}_2$
 with state $(\mathbf{x})^t = \underline{x}^t = \mathcal{A}^t(\underline{x}^0)$
 reached at from $\underline{x}^0$ after $t\in \mathbb{N}$ time steps. $\mathbb{N}$ denotes the set of nonnegative integers.

 One can see that the time, space, and states of this system take only discrete values. The first ECA that we consider here evolves according to the local rules
 %..................................
 \begin{align} %\label{local rule1}
 x_{i}^{t+1} &\equiv \mathcal{A}_{90} (x_{i-1}^t, x_{i}^t, x_{i+1}^t) = [x_{i-1}^t + x_{i+1}^t] {\rm mod}\ 2~, \nonumber \\
 x_{i}^{t+1}&\equiv \mathcal{A}_{165} (x_{i-1}^t, x_{i}^t, x_{i+1}^t) =1-[x_{i-1}^t + x_{i+1}^t] \rm{mod}\ 2~, \nonumber
 \end{align}
 which correspond to the rule 90 and its complementary rule 165 described in the tabular form below:
% The following is the lookup table of rule 90.

  \medskip

    \begin{center}
    \begin{tabular}{lllllllll}
        \hline
        %\multicolumn{1}{|l|}{Number} &
%        \multicolumn{1}{|c|}{7} &\multicolumn{1}{|c|}{6} & \multicolumn{1}{|c|}{5} & \multicolumn{1}{|c|}{4} &
%        \multicolumn{1}{|c|}{3} & \multicolumn{1}{|c|}{2} & \multicolumn{1}{|c|}{1} & \multicolumn{1}{|c|}{0} \\
%        \hline
        \multicolumn{1}{|l|}{Current neighbourhood pattern:} &
        \multicolumn{1}{|c|}{111} & \multicolumn{1}{c|}{110} & \multicolumn{1}{c|}{101} & \multicolumn{1}{c|}{100} & \multicolumn{1}{c|}{011} & \multicolumn{1}{c|}{010} & \multicolumn{1}{c|}{001} & \multicolumn{1}{c|}{000} \\
        \hline
        \multicolumn{1}{|l|}{Rule result for central cell:$\quad$ $^{90}_{165}$} &
        \multicolumn{1}{|c|} {$^{0}_{1}$} & \multicolumn{1}{c|}{$^{1}_{0}$} & \multicolumn{1}{c|}{$^{0}_{1}$} & \multicolumn{1}{c|}{$^{1}_{0}$} & \multicolumn{1}{c|}{$^{1}_{0}$} & \multicolumn{1}{c|}{$^{0}_{1}$} & \multicolumn{1}{c|}{$^{1}_{0}$} & \multicolumn{1}{c|}{$^{0}_{1}$} \\
        \hline
    \end{tabular}
    \end{center}

 \medskip

 The second row shows the future state of the cell if the cell itself and its left and right neighbors are in the arrangement shown in the first row.
    In fact, a rule is numbered by the unsigned decimal equivalent of the binary expression in the second row.
    For the second complementary pair considered here, the local rules are:
 %..................................
 \begin{align} %\label{local rule2}
 x_{i}^{t+1} &\equiv \mathcal{A}_{150} (x_{i-1}^t, x_{i}^t, x_{i+1}^t) = [x_{i-1}^t + x_{i}^t+x_{i+1}^t] {\rm mod}\ 2~, \nonumber\\
 x_{i}^{t+1} &\equiv \mathcal{A}_{105} (x_{i-1}^t, x_{i}^t, x_{i+1}^t)=1-[x_{i-1}^t + x_{i}^t+ x_{i+1}^t] \rm{mod}\ 2~,\nonumber
 \end{align}
whereas their table form is the following one:
    \begin{center}
   \begin{tabular}{lllllllll}
       \hline
       \multicolumn{1}{|l|}{Current neighbourhood pattern:} &
       \multicolumn{1}{|c|}{111} & \multicolumn{1}{c|}{110} & \multicolumn{1}{c|}{101} & \multicolumn{1}{c|}{100} & \multicolumn{1}{c|}{011} & \multicolumn{1}{c|}{010} & \multicolumn{1}{c|}{001} & \multicolumn{1}{c|}{000} \\
       \hline
       \multicolumn{1}{|l|}{Rule result for central cell: \, $\displaystyle {^{150}_{105}}$}
       & \multicolumn{1}{|c|} {$^{1}_{0}$} & \multicolumn{1}{c|}{$^{0}_{1}$} & \multicolumn{1}{c|}{$^{0}_{1}$} & \multicolumn{1}{c|}{$^{1}_{0}$} & \multicolumn{1}{c|}{$^{0}_{1}$} & \multicolumn{1}{c|}{$^{1}_{0}$} & \multicolumn{1}{c|}{$^{1}_{0}$} & \multicolumn{1}{c|}{$^{0}_{1}$} \\
       \hline
   \end{tabular}
   \end{center}
 \medskip

The uniform ECA are those for which the same rule is applied to update the cells as in this paper, otherwise the ECA are called non-uniform or hybrid.
%It is also worth mentioning that the evolution rules of ECA are determined by two main factors: the rule and the initial conditions.

The time series that we consider in the case of the four ECA rules examined here belong to the row sum signals as used by Nagler and Claussen \cite{NC}, i.e., the sums of the 1's in each row (at each time step). Examples of the corresponding time series for the four ECA rules considered in this paper are displayed in Fig.~1.

It is also worth mentioning that the evolution rules of ECA are determined by two main factors: the rule and the initial conditions.
For example, for our time series samples, in the case of the rule 90 with a single central 1 in the first row we have the first peak of amplitude $2^{11}$ at time (row) $2^{11}$, while the same peak is shifted two rows upwards if the initial condition is three 1's in the central position.
On the other hand, for the rule 150 with a central 1 in the first row there are no peaks of height $2^{11}$ but instead we have found peaks of height
$2^{11}+1$ at the rows 1536, 2046, 2560, and so forth.

%and their multifractal singularity spectra determined with the three algorithms of this paper are displayed in Figs. 1 and 2.

%%%%% บบบบบบบบบบบบบบบบบบบบบบบบบบบบบบบบบบบบบบบบบบบบFIG. 1
\begin{figure}[h!] %[h!]
 \centering
 \includegraphics[width=9.5cm, height=11cm]{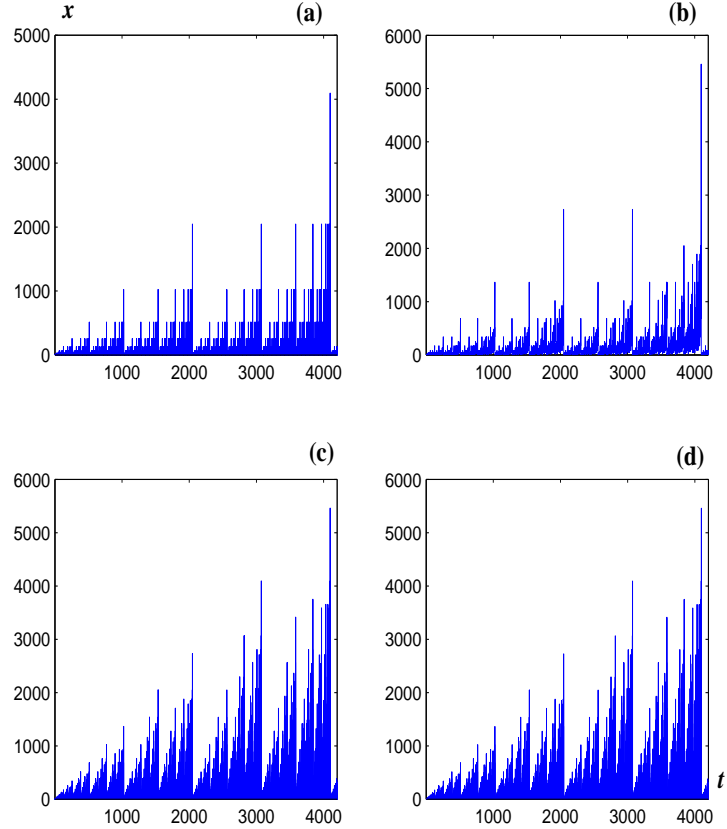}  %{fig_AMF_R90_165_SSpectrums2_13vF.eps}
  \caption{\sl \small
  Time series of the row sum signal with a single 1 in the central position of the first row as initial condition corresponding to:
  (a) rule R90, (b) its complementary rule R165, (c) rule R150, and (d) its complementary rule R105.
  Only the first $2^{12}$ points are shown of the whole set of $2^{16}$ data points. Since the mean signals for the complementary rules are much higher than for the rules themselves showing also some specific trends we eliminated their contribution in the time series of the right column.
  %In (c) and (d) are the respective singularity spectrums $f(\alpha)$ of the time series of the row signal.
 }
 \label{fig-90-165}
\end{figure}
%%%%%% บบบบบบบบบบบบบบบบบบบบบบบบบบบบบบบบบบบบบบบบบบบบFIG. 2
%\begin{figure}[h!] %[h!]
% \centering
% \includegraphics[width=9.5cm, height=11cm]{fig_AMF_R150_105_SSpectrums2_13vF.eps}
%  \caption{\sl \small
%  Time series of the row signal corresponding to rule (a) R150, and its complementary
%  rule (b) R105.
%  Only the first $2^{9}$
%  points are shown of the whole set of $2^{16}$ data points.
%  In (c) and (d) are the respective singularity spectrums $f(\alpha)$ of the time
%  series of the row signal.
% }
% \label{fig-150-105}
%\end{figure}

\section{Brief Description of the Multifractal Methods}

In a previous work \cite{epl09}, we used the wavelet-transform detrended fluctuation analysis proposed by Manimaran \cite{mani}, henceforth WT-MFDFA, to perform the multifractal analysis of three relevant ECA time series and found multifractal spectra estimates in quantitative agreement with the ones obtained with the multifractal detrended fluctuation analysis (MFDFA \cite{K03}). However, recently, the literature abounds in comparison works related to the existing numerical methods for multifractal behaviour, which in fact try to establish their advantages and disadvantages one with respect to the others \cite{H11}. According to those works, the so-called wavelet leaders (WL), a scaling method that employs the coefficients of an orthogonal wavelet decomposition of the signal, seems to perform better than other competing methods, such as the WT modulus maxima (WTMM) and the MFDFA. O\'swi\c ecimka et al. \cite{Osw} processed  synthesized data with WTMM and MFDFA and stated that the MFDFA provides a better estimation of singularity spectrum than WTMM. Jaffard et al. \cite{J05} claimed that also WL provides a better singularity spectrum than WTMM. In addition, a comparison study of Serrano and Figliola \cite{SF09} between WL and MFDFA favoured WL, although for  short time series they recommend MFDFA to extract the multifractal spectrum.
Regarding the estimation of the Hurst exponent, we note that in the review paper of Mielniczuk and Wojdyllo \cite{MW06} it is claimed that the properties of several variants of the DFA estimators differ greatly in their performance!

%%% ============================
 \subsection{WT-MFDFA}\label{s-DFA-WT}
% {\bf WMF-DFA}. -
%%% ============================
An important class of wavelets are those with the vanishing moment property (see the Appendix) which directly helps to detrend the data.
To reveal the MF properties %\cite{Halsey}
of ECA one should separate the trend from fluctuations in the ECA time series. We decided to use the discrete wavelet method proposed by Manimaran {\em et al.} \cite{mani}. This method exploits the fact that the low-pass version resembles the original data in an ``averaged'' manner in different resolutions. Instead of a polynomial fit, we consider the different versions of the low-pass coefficients to calculate the ``local'' trend. Let $x(t_k)$ be a time series type of data, where $t_k=k \Delta t$ and $k=1,2,\ldots,~N$. %=2^j$.
Then the algorithm that we employ contains the following steps:

  \begin{enumerate}

\item  Determine the profile $Y(k)$ of the time series, which is the cumulative sum of the series from which the series mean value is subtracted, i.e., $Y(k)=\sum_{i=1}^{k} (x(t_i)-\langle x\rangle )$.

    \item Calculate the fast wavelet transform (FWT), i.e., the multilevel wavelet decomposition of the profile.
          For each level $m$, we get the fluctuations of the $Y(k)$ by subtracting the ``local''
          trend of the $Y$ data, i.e., $ \Delta Y(k;m) = Y(k) - \tilde{Y}(k;m)$, where $\tilde{Y}(k;m)$ is the reconstructed profile after removal of successive details coefficients at each level $m$. These fluctuations at level $m$ are subdivided into windows, i.e., into $M_s={\rm int}(N/s)$ non-overlapping segments of length $s$. This division is performed starting from both the beginning and the end of the fluctuations series (i.e., one has $2M_s$ segments). Next, one calculates the local variances associated to each window $\nu$ %= 1,..., M_s$
%..............
         \begin{equation}\label{eq-Fs1}
           F^2(\nu,s;m)={\rm var}\Delta Y((\nu-1)s+j;m)~, \quad j=1,..., s~, \quad \nu=1,..., 2M_s~. %\quad M_s={\rm int}(N/s)~.
         \end{equation}

    \item  Calculate a $q$th order fluctuation function defined as
%........................................
          \begin{equation}\label{eq-Fqs}
            F_q(s;m) = \left\{ \frac{1}{2M_s} \sum_{\nu=1}^{2M_s} |F^2(\nu,s;m)|^{q/2} \right\}^{1/q}
          \end{equation}

          where $q \in \mathbb{Z}$ with $q \neq 0$. The
          diverging behaviour for $q\to 0$ can be avoided by using
          a logarithmic averaging $F_0(s;m) = \exp\left\{\frac{1}{2M_s} \sum_{\nu=1}^{2M_s} \ln |F^2(\nu,s;m)| %\frac{1}{2N} \sum_{k=1}^{N} \ln |F(m, k)|
          \right\}$, see \cite{K02}.
  \end{enumerate}

    If the fluctuation function $F_q(s;m)$ displays a power law scaling
%..............................
       \begin{equation}\label{eq-FqsPLaw}
         F_q(s;m) \sim s^{h(q)},
       \end{equation}
%.............................
then the analyzed time series has a fractal scaling behaviour. The exponent $h(q)$ is the generalized Hurst exponent since it depends on $q$, while the original Hurst exponent is $h(2)$. If $h(q)$ is constant for all $q$ then the time series is monofractal, otherwise it has a MF behavior. In the latter case, one can calculate various other MF scaling exponents, such as $\tau(q)$ and $f(\alpha)$ \cite{Halsey}.

\subsection{MFDFA}

MFDFA is an algorithm developed in 2002 by Kantelhardt et al \cite{K02}, which differ from WT-MFDFA by the fact that it does not make use of WT for detrending. In fact the origin of both procedures is the conventional detrended fluctuation analysis of Peng et al \cite{Pengetal}. Again the data to be considered are profiles of the signals, which are divided into non-overlapping segments both from the beginning and from the end of the set of data getting therefore $2N_s$ segments altogether. In each of these segments the local trend is calculated by a least square fit of the series.
Then one calculates the variance $F^2(s,\nu)$ for each segment $\nu$ with respect to the fitting line $y_\nu$ in each segment $\nu$.
Finally, an average over all segments is performed leading to the $q$--th order fluctuation function similar to (\ref{eq-Fqs})
 \begin{equation}\label{eq-FqsK02}
            F_q(s) = \left\{ \frac{1}{2N_s} \sum_{\nu=1}^{2N_s} |F^2(\nu,s)|^{q/2} \right\}^{1/q}~.
          \end{equation}

 \subsection{WTMM}

In the presence of an isolated singularity in the data at a particular point $t_0$, the scaling behavior of the wavelet coefficients (see equation (\ref{eq-CWT}) in the appendix) is described by the H\"older exponent $\alpha(t_0)$ as

%%%%% ===========
\begin{equation} \label{eq-singular}
 W_x(a, t_0) \sim a^{\alpha(t_0) + 1/2},
\end{equation}
%%%%% ===========
%%%%% ===========
in the limit $a \to 0^+$. To characterize the singular behavior of functions, it is sufficient to consider the values and positions of the Wavelet Transform Modulus  Maxima (WTMM) \cite{mallat-huang} defined as points $(a_0, b_0)$ on the scale-position plane, $(a,b)$, where $|W_x(a_0, b)|$ is locally maximum for $b$ in the neighborhood of $b_0$. These maxima are located along curves in the plane $(a,b)$.
 However, the relationship \eqref{eq-singular} in some cases is not appropriated to describe distribution functions with non-isolated singularities.
 The wavelet multifractal formalism may characterize fractal objects which cannot be completely described using a single fractal dimension.
 According to Bacry {\em et al.} \cite{arne1}, an ``optimal'' partition function ${\mathcal Z}_q(x, a)$ can be defined in terms of the WTMM.
 They consider the set of modulus maxima at a scale $a$ as a covering of the singular
 support of $x$ with wavelets of scale $a$. The partition function ${\mathcal Z}_q$ measures the sum of all wavelet modulus
 maxima at a power $q$ as follows
%%%%% ===========
\begin{equation} \label{Mallat:FP}
 {\mathcal Z}_q(x, a) = \sum_p |W_x(a, b_p(a)) |^q,
\end{equation}
%%%%% ===========
where $\{ b_p(a) \}_{p \in \mathbb{Z}}$ is the position of all local maxima of $|W_x(a, b) |$ at a fixed scale $a$. This partition function
is very close to the definition of the partition function introduced in \cite{Halsey}.
It can be inferred from \eqref{Mallat:FP} that for $q>0$ the most pronounced modulus maxima will prevail, whereas for $q<0$ the lower ones will survive. For each $q \in \mathbb{R}$, the partition function %${\mathcal Z}_q(f, a)$
 is related to its scaling exponent $\tau(q)$ in the following way ${\mathcal Z}_q(x, a) \sim a^{\tau(q)}$.
 A linear behavior of $\tau(q)$ indicates monofractality whereas nonlinear behavior suggests that a signal is a
 multifractal. A fundamental result in the (wavelet) multifractal formalism states that the singularity (H\"older) spectrum $f(\alpha)$ of
 the signal $x(t)$ is the Legendre transform of $\tau(q)$, i.e.,
%%%%% ===========
\begin{equation} \label{Leg:D-tau}
    \alpha = \frac{d\tau(q)}{dq}, \qquad \text{and} \qquad  f(\alpha)  = q\alpha - \tau(q).
\end{equation}
%%%%% ===========
The H\"older spectrum of dimensions $f(\alpha)$ is a non-negative convex function that is supported on
the closed interval $[\alpha_{\text{min}}, \alpha_{\text{max}}]$, which is interpreted as the Hausdorff fractal dimension of
 the subset of data characterized by the H\"older exponent $\alpha$ \cite{Salo1}.
The most ``frequent'' singularity, which corresponds to the maximum  of $f(\alpha)$, occurs for the value of $\alpha(q=0)$, whereas the boundary
values of the support, $\alpha_{\text{min}}$ for $q>0$ and $\alpha_{\text{max}}$ for $q<0$, correspond to the
 strongest and weakest singularity, respectively.

 The analyzing wavelets which are used most frequently are the successive derivatives of the Gaussian function
%.........................
  \begin{equation}\label{eq-Wavelets}
      \psi^{(n)}(t) := \frac{d^n}{dt^n}\left( \exp(-t^2 / 2)\right), \qquad n \in
      \mathbb{Z}^+,
  \end{equation}
 because they are well localized both in space and frequency, and they remove the trends of the signal that can be approximated by polynomials
 up to $(n- 1)$th order. %\cite{arne3}.
 In particular, our analyses were carried out with the Mexican hat wavelet $\psi^{(2)}(t)$.

\section{Data analyses}

\begin{figure}[h!] %[h!]
 \centering
 \includegraphics[width=12.5cm, height=5cm]{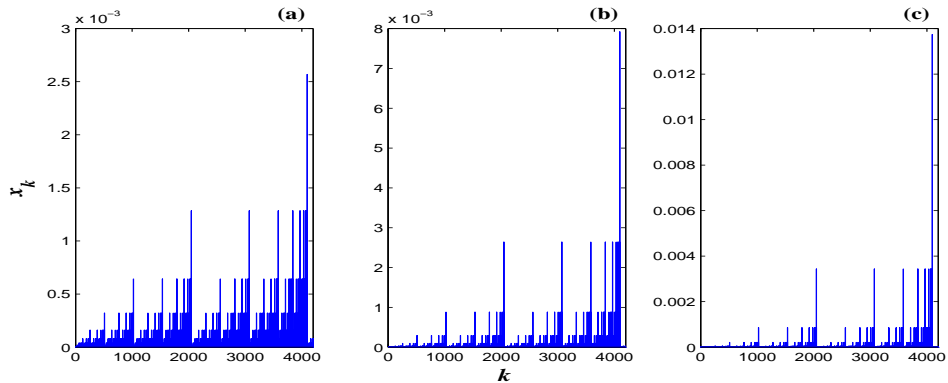}
  \caption{\sl \small Binomial time series $x_k$ for the binomial model (\ref{eq-Binomial}) of parameters $c=2/3$, $3/4$, and $4/5$ from left to right, respectively. We have used $n_{\rm max}=13$ but only the first $2^{12}$ data are shown.
 }
 \label{fig-BinomTS}
\end{figure}

In order to illustrate the efficiency and the fitting properties of the three procedures, we first carry out the analysis of the binomial multifractal time series, whose scaling parameters are analytically known \cite{K02,Feder}. The binomial multifractal time series are series of $N = 2^{n_{\textrm{max}}}$ numbers $x_k$, with $k = 1, \ldots , N$, defined by
%.......................
 \begin{equation}\label{eq-Binomial}
            x_k = c^{n(k-1)}(1-c)^{n_{\textrm{max}}- n(k-1)}.
        \end{equation}
where $0.5 < c < 1$ is a parameter and $n(k)$ is the number of digits equal to 1 in the binary representation of the index $k$. Examples of binomial time series $x_k$ for three values of the parameter $c$ are given in Fig.~(\ref{fig-BinomTS}).

        %%%%% บบบบบบบบบบบบบบบบบบบบบบบบบบบบบบบบบบบบบบบบบบบบFIG. 2

The binomial multifractal exponents $h(q)$ and $\tau(q)$ can be calculated in closed forms
%..................................
        \begin{equation}\label{eq-Scaling-Exp}
            h(q) = \frac{1}{q} - \frac{\ln [c^q + (1-c)^q]}{q \ln 2}, \qquad
            \tau(q) = - \frac{\ln [c^q + (1-c)^q]}{\ln 2}.
        \end{equation}

\begin{figure}[h!] %[h!]
 \centering
 \includegraphics[width=7.5cm, height=9cm]{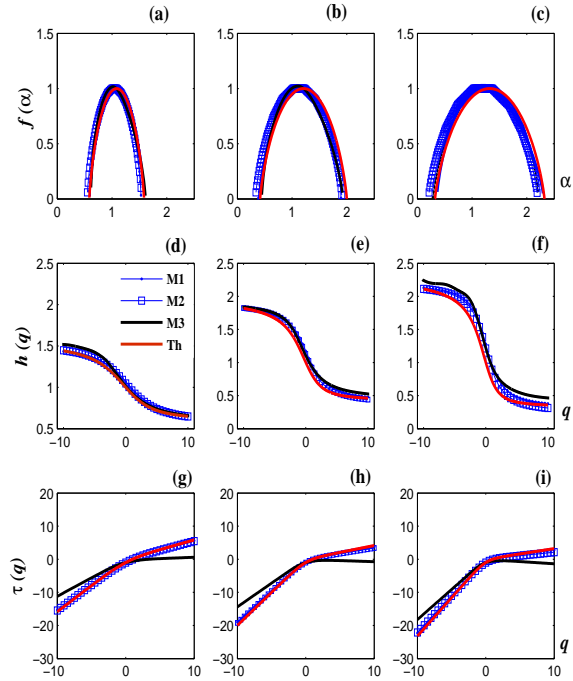}    %{fig_AMF_BinomialModel_VF.eps}
  \caption{\sl \small
  The singularity spectrum $f(\alpha)$, the generalized Hurst exponent, and the $\tau$ exponent for the binomial multifractal model of parameters
  $c=2/3$ (left column), $c=3/4$ (middle column), and $c=4/5$ (right column). $\rm M1$ corresponds to WT-MFDFA, $\rm M2$ to MFDFA, and $\rm M3$ to WTMM, while $\rm Th$ corresponds to the theoretical binomial formulas for the same parameters.
 }
 \label{fig-BinomialM}
\end{figure}

         %Table \ref{tab-Hq} and
        Figure~\ref{fig-BinomialM} %- \ref{fig_AMF_R150_105_SSpectrums2_13vF.eps}
        shows the multifractal singularity spectrum $f(\alpha)$, the generalized Hurst exponent, and the $\tau$ exponent for three values of the binomial parameter $c$.
        %between the values for the theoretical case ($h_T(q)$), with the
%        numerical results obtained through wavelet analysis ($h_W(q)$).

        %%%%% บบบบบบบบบบบบบบบบบบบบบบบบบบบบบบบบบบบบบบบบบบบบFIG. 3

A direct inspection of the binomial multifractal plots shows that the numerical differences between the three methods are not significant in the case of the singularity spectra and the generalized Hurst exponent, although the WT-MFDFA and MFDFA are closer to the theoretical curves than the WTMM method. In the case of the $\tau$ exponent, the WTMM clearly indicates a more pronounced multifractality than the theoretical one, while the first two methods are again very close to the theoretical curve. We have checked that the high $|q|$ discrepancies of the WTMM $\tau$ exponent stay almost constant at least for sample sizes up to $2^{16}$ data but we do not rule out the WTMM procedure because as we said the singularity spectra and the Hurst exponent  obtained by this method do not appear to be different from the binomial theoretical results in a relevant manner.

We proceed with the comparison study among the three scaling approaches, namely the WT-MFDFA, MFDFA, and WTMM analyses, for the time series of the row sum signals of two pairs of complementary ECA rules (90, 165) and (105,150) with the goal of providing further evidence of their multifractal nature and also to check the performance of the three scaling methods. The idea behind the latter test is simple. Since the rules in the pairs are complementary in the sense that their transition rates fulfill $f_{105}=1- f_{150}$ and $f_{90}=1- f_{165}$ they should not differ statistically for samples of reasonable sizes and should provide quite similar results. Therefore, the most reliable scaling method is the one which gives the smallest fluctuations in the differences of the scaling parameters between the two members of any complementary ECA pair.

\bigskip

\begin{figure}[h!] %[h!]
 \centering
 \includegraphics[width=7.5cm, height=9cm]{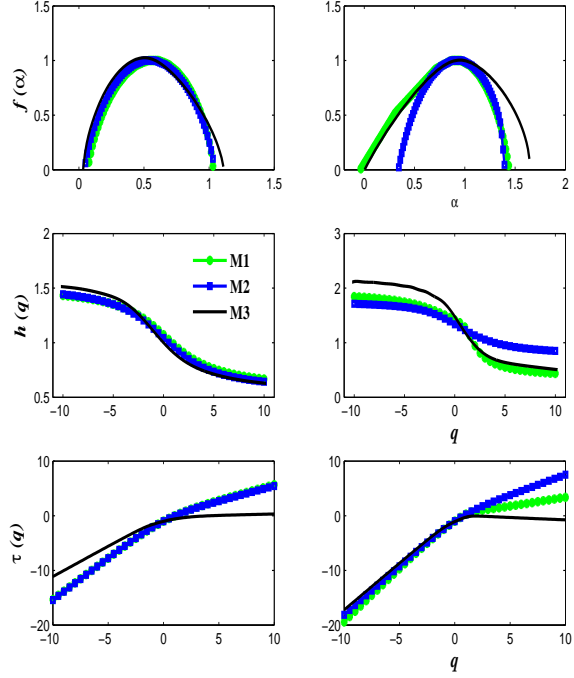}
  \caption{\sl \small
   The singularity spectrum $f(\alpha)$, the generalized Hurst exponent, and the $\tau$ exponent for the rule pair (90--165) for the three multifractal methods employed here.
 }
 \label{fig-1st Pair}
\end{figure}

\begin{figure}[h!] %[h!]
 \centering
 \includegraphics[width=7.5cm, height=9cm]{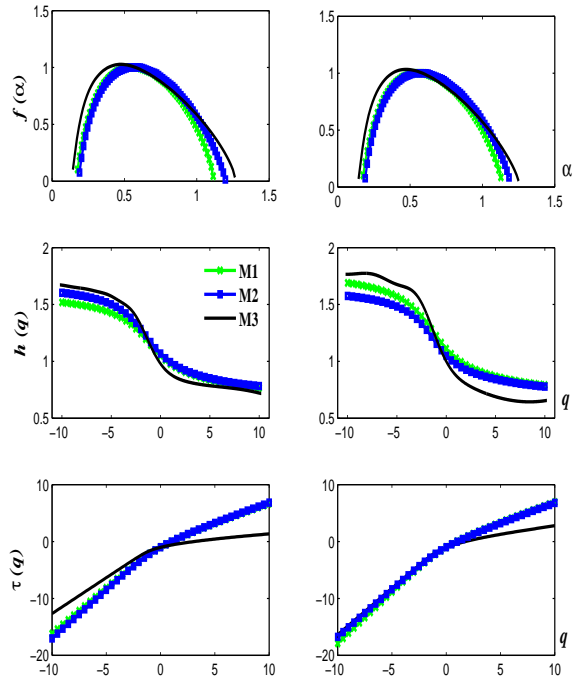}
  \caption{\sl \small
  The singularity spectrum $f(\alpha)$, the generalized Hurst exponent, and the $\tau$ exponent for the rule pair (150--105) for the three multifractal methods as above.
 }
 \label{fig-2nd Pair}
\end{figure}

The main multifractal quantities are displayed in Fig.~\ref{fig-1st Pair} for the first ECA pair and in Fig.~\ref{fig-2nd Pair} for the second one. As in the binomial case, the simple inspection of the plots indicates the following features. WT-MFDFA is closer to WTMM on the left wing of the singularity parabola ($\alpha < \alpha_{\rm max}$), whereas it is closer to the MFDFA on the right wing ($\alpha > \alpha_{\rm max}$). In the case of $h(q)$, WT-MFDFA is closer to MFDFA at negative $q$s but closer to WTMM at positive $q$s. Finally, for $\tau(q)$ WT-MFDFA is always closer to MFDFA; in fact, for rule 90 the results of the two methods are almost identical. At the same time, WT-MFDFA indicates a less pronounced difference between the fractal components than MFDFA and WTMM, as taken in this order. Similar results are obtained for the second pair although these tendencies are less pronounced.

\bigskip

\begin{figure}[x] %[h!]
 \centering
  \includegraphics[width=11.5cm, height=13cm]  {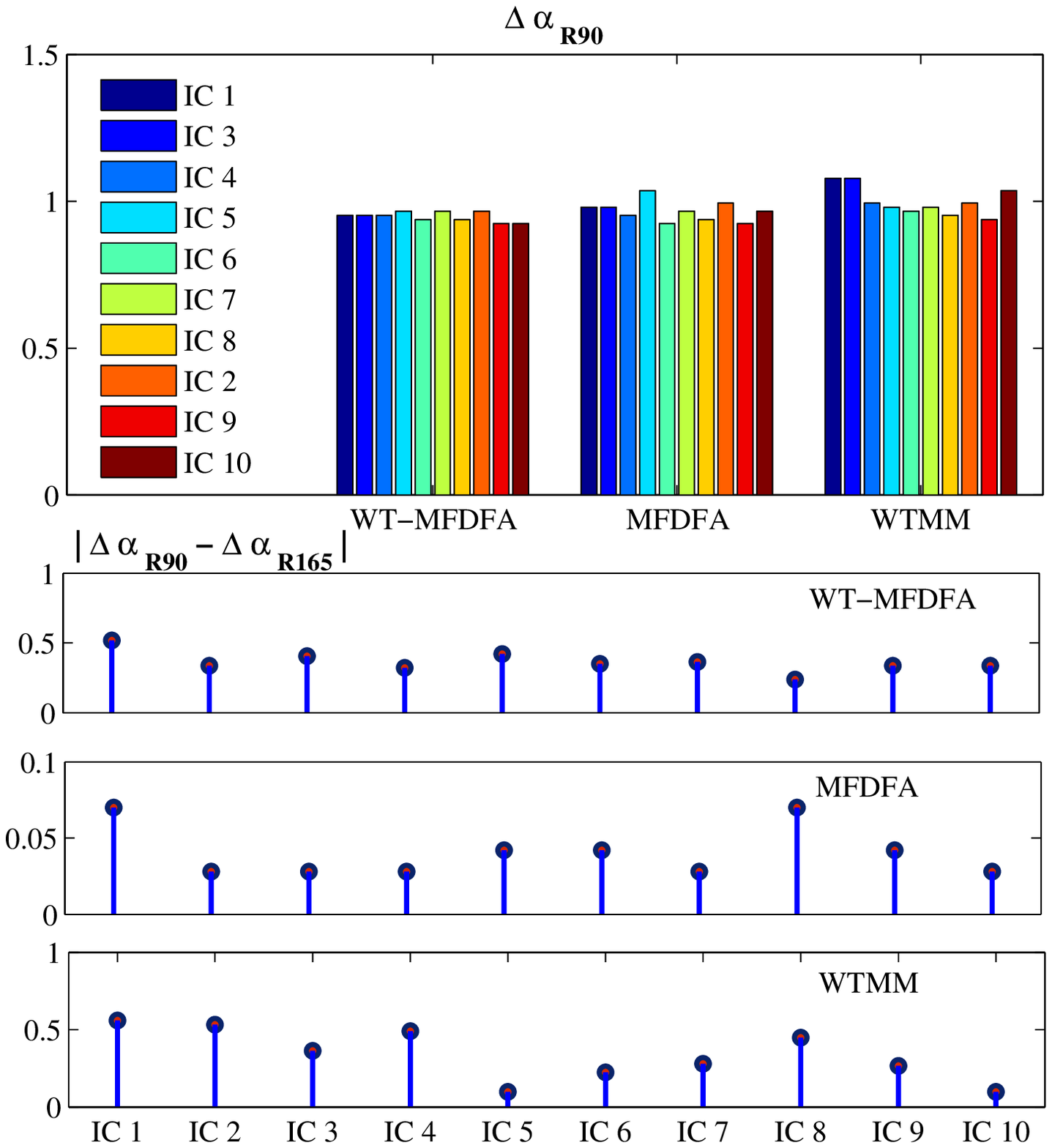}   %{13f-R90_Db2.eps}   %{R90NA_db6_db16}
   \caption{\sl \small
   Rule 90 ECA: The multifractal support $\Delta \alpha$ in the case of the row sum signal of a sequence of $2^{13}$ time steps with the ten initial conditions as explained in the text. There are no significant differences in the case of the three scaling methods that we tested. In the lower three plots, the differences $|\Delta \alpha_{R90}-\Delta \alpha_{R165}|$ between the multifractal supports of the two complementary rules for each scaling methods are displayed. The MFDFA procedure gives differences less than 0.08 for the ten tested initial conditions, much better than WT-MFDFA and WTMM.
   %Rule 90: (a) A part of the row signal selected from a total sample of $2^{13}$ time steps. (b) The maxima. (c) The $\tau$ exponent, $\tau(q)$. (d) % The singularity spectrum $f(\alpha)=q\frac{d\tau(q)}{dq}-\tau(q)$ of the row signal.
   %The calculations of the multifractal quantities $h$, $\tau$, and $f(\alpha)$ are performed both with MF-DFA and the wavelet-based WMF-DFA.
   %Non-cumulative time series and MF quantities for rule 90. For negative $q$s we used db6 wavelets whereas for positive $q$s we used db16 ones.
  }
  \label{fig-C1}
\end{figure}

We merely use the WT-MFDFA here rather than WL because we employed it in our previous paper \cite{epl09} and in fact both procedures are based on the coefficients of an orthogonal wavelet decomposition. Our results for the extension (support) of the multifractality, the most frequent H\"older exponent, and the Hurst exponent are presented graphically in Figures~\ref{fig-C1}, \ref{fig-C3}, and \ref{fig-C5}, respectively, for the ECA pair (90,165) and in Figures~\ref{fig-C2}, \ref{fig-C4}, and \ref{fig-C6}, for the pair (150,105) as three groups of ten histograms corresponding to the three methods and ten initial conditions of centered activity type, i.e., with symmetrically distributed 1's around the central position in the first row. As commented in the captions, the plots indicate that in the case of the multifractal support, the MFDFA gives smaller differences between the two complementary rules than WT-MFDFA and WTMM, while for the other two scaling parameters the WT-MFDFA is the best from this point of view. We also mention that the arguments of Huang and collaborators \cite{H11} against representing a continuous signal through a discrete WT do not apply to our data because the CA signals are naturally discrete.
%....................... FIG 4

%...................... FIG 5

  \section{Conclusions}

In conclusion, we provided here a strong confirmation of our previous results \cite{epl09} that the implementation of the discrete WT in the MFDFA method leads to the WT-MFDFA as a very acceptable algorithm to obtain the multifractal ECA parameters with much less computational cost and better accuracy. We also found that there is no apparent trend in the results depending on an extended set of initial conditions represented by an extended centered initial activity in the CA terminology. Since we used complementary automaton rules, which are statistically identical we could compare the numerical results of the WT-MFDFA, MFDFA, and WTMM procedures. According to the criterion of minimum differences in the scaling quantities for statistically identical data, it appears that the MFDFA performs better than WT-MFDFA and WTMM in the case of the multifractal support, while for the most frequent H\"older exponent and the Hurst exponent the WT-MFDFA is the best. The same numerical techniques have been tested for the analytically known case of binomial multifractal time series for three binomial parameters with the result that WT-MFDFA and MFDFA reproduce very closely the theoretical values of the multifractal quantities. This can be interpreting as compelling evidence that the latter numerical methods are reliable tools to calculate the multifractal properties.

                       %%%%% บบบบบบบบบบบบบบบบบบบบบบบบบบบบบบบบบบบบบบบบบบบบFIG. C1
%%%%% บบบบบบบบบบบบบบบบบบบบบบบบบบบบบบบบบบบบบบบบบบบบFIG. C2
 \begin{figure} [x] \label{fig-C2}%[h!]
 \centering
  \includegraphics[width=11.5cm, height=13cm]  {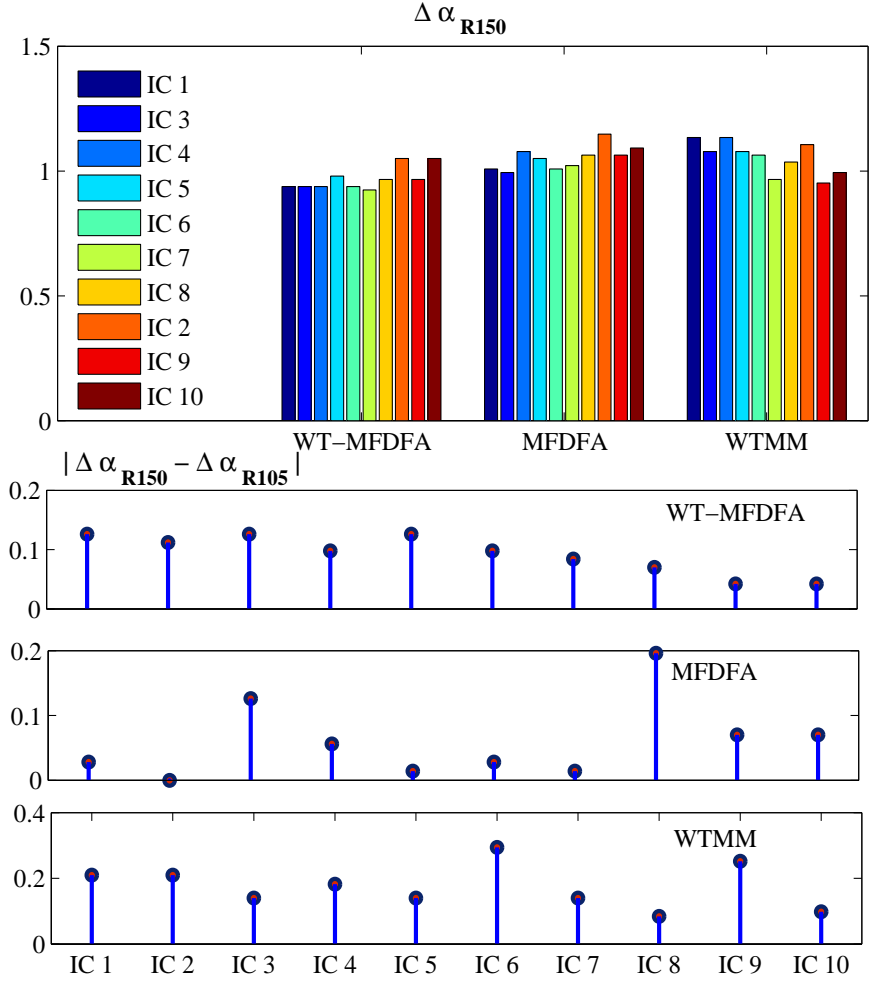}   %{13f-R90_Db2.eps}   %{R90NA_db6_db16}
   \caption{\sl \small
   Rule 150 ECA: The multifractal support $\Delta \alpha$ in the case of the row sum signal of a sequence of $2^{13}$ time steps with the same initial conditions. There are no significant differences in the case of the three scaling methods that we tested. In the lower three plots, the differences $|\Delta \alpha_{R150}-\Delta \alpha_{R105}|$ between the multifractal supports of the two complementary rules for each scaling methods are displayed. Although WT-MFA and WTMM performs better for this complementary pair than for the other one, the MFDFA procedure still provides less differences.
   %(b) The maxima. (c) The $\tau$ exponent, $\tau(q)$. (d) % The singularity spectrum $f(\alpha)=q\frac{d\tau(q)}{dq}-\tau(q)$ of the row signal.
   %The calculations of the multifractal quantities $h$, $\tau$, and $f(\alpha)$ are performed both with MF-DFA and the wavelet-based WMF-DFA.
   %Non-cumulative time series and MF quantities for rule 90. For negative $q$s we used db6 wavelets whereas for positive $q$s we used db16 ones.
  }
  \label{fig-C2}
\end{figure}
%%%%% บบบบบบบบบบบบบบบบบบบบบบบบบบบบบบบบบบบบบบบบบบบบFIG. C3
 \begin{figure} [x] %[h!]
 \centering
  \includegraphics[width=11.5cm, height=13cm]  {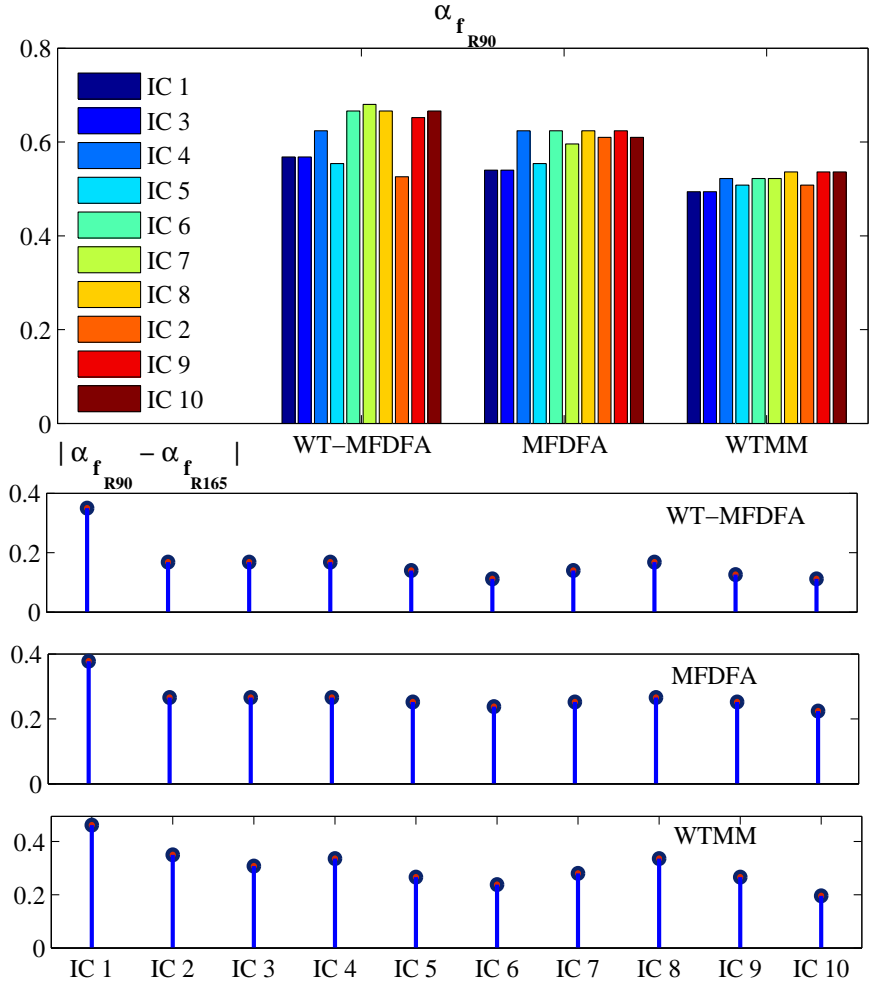}   %{13f-R90_Db2.eps}   %{R90NA_db6_db16}
   \caption{\sl \small
   Rule 90 ECA: The most frequent H\"older exponent $\alpha_f$ obtained with WT-MFA, MFDFA, and WTMM for the ten initial conditions. The lower three plots display the difference $|\alpha_{f_{R90}}-\alpha_{f_{R165}}|$. In this case, the WT-MFDFA works better.
   %(a) A part of the row signal selected from a total sample of $2^{13}$ time steps. (b) The maxima. (c) The $\tau$ exponent, $\tau(q)$. (d) % The singularity spectrum $f(\alpha)=q\frac{d\tau(q)}{dq}-\tau(q)$ of the row signal.
   %The calculations of the multifractal quantities $h$, $\tau$, and $f(\alpha)$ are performed both with MF-DFA and the wavelet-based WMF-DFA.
   %Non-cumulative time series and MF quantities for rule 90. For negative $q$s we used db6 wavelets whereas for positive $q$s we used db16 ones.
  }
 \label{fig-C3}
\end{figure}
%%%%% บบบบบบบบบบบบบบบบบบบบบบบบบบบบบบบบบบบบบบบบบบบบFIG. C4
\begin{figure} [x] %[h!]
 \centering
  \includegraphics[width=11.5cm, height=13cm]  {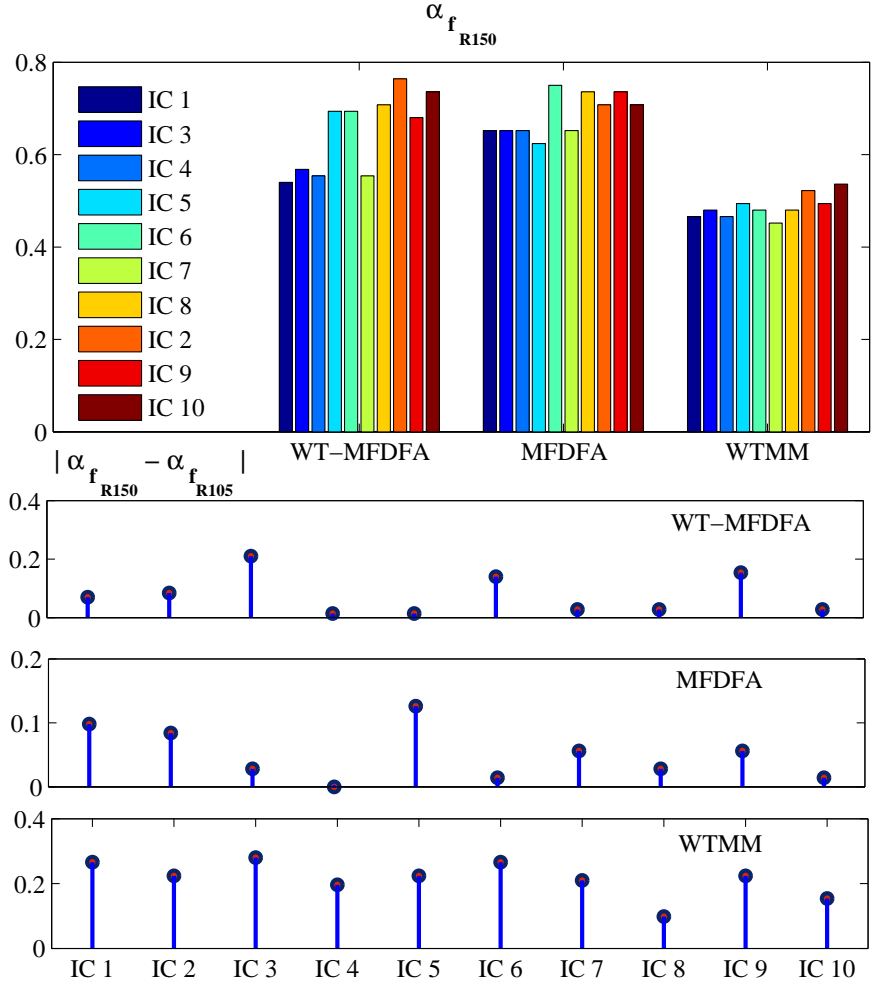}   %{13f-R90_Db2.eps}   %{R90NA_db6_db16}
   \caption{\sl \small
   Rule 150 ECA: The most frequent H\"older exponent $\alpha_f$ obtained with WT-MFA, MFDFA, and WTMM for the same initial conditions. The lower three plots display the difference $|\alpha_{f_{R150}}-\alpha_{f_{R105}}|$. The WT-MFDFA works better again.
   %Rule 90: (a) A part of the row signal selected from a total sample of $2^{13}$ time steps. (b) The maxima. (c) The $\tau$ exponent, $\tau(q)$. (d) % The singularity spectrum $f(\alpha)=q\frac{d\tau(q)}{dq}-\tau(q)$ of the row signal.
   %The calculations of the multifractal quantities $h$, $\tau$, and $f(\alpha)$ are performed both with MF-DFA and the wavelet-based WMF-DFA.
   %Non-cumulative time series and MF quantities for rule 90. For negative $q$s we used db6 wavelets whereas for positive $q$s we used db16 ones.
  }
 \label{fig-C4}
\end{figure}
%%%%% บบบบบบบบบบบบบบบบบบบบบบบบบบบบบบบบบบบบบบบบบบบบFIG. C5
\begin{figure} [x] %[h!]
 \centering
  \includegraphics[width=11.5cm, height=13cm]  {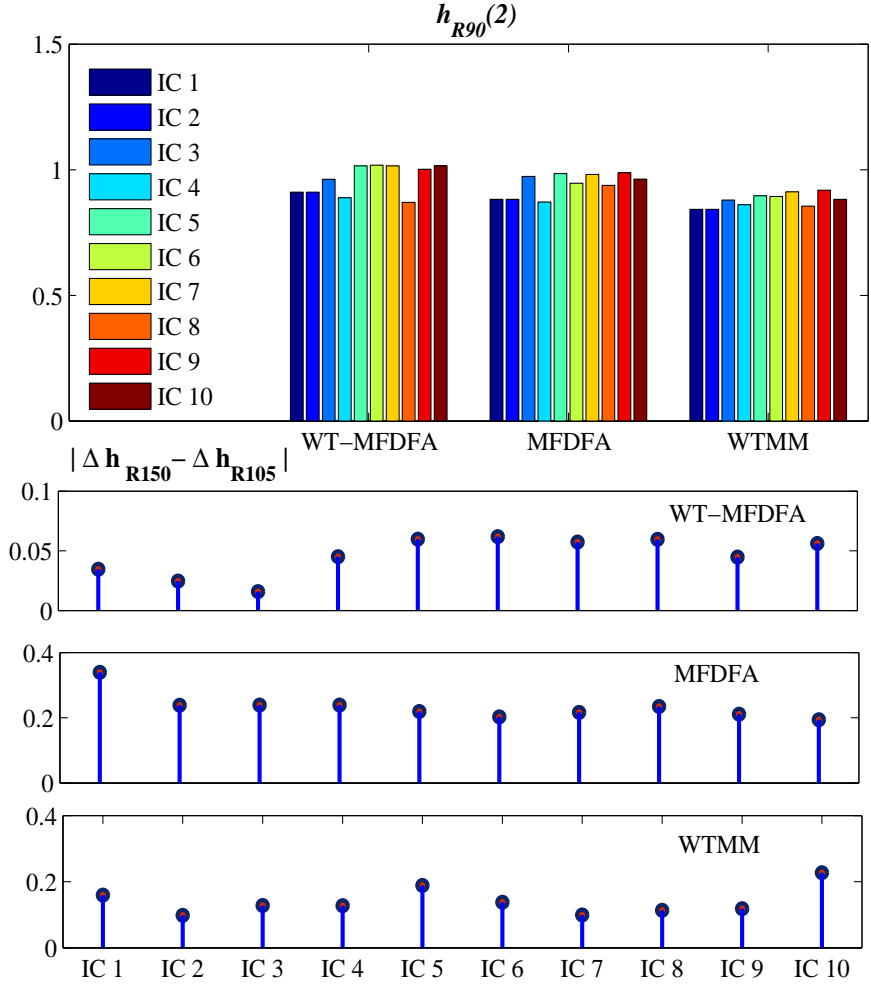}   %{13f-R90_Db2.eps}   %{R90NA_db6_db16}
   \caption{\sl \small
   Rule 90 ECA:  The Hurst exponent $h_{R90}(2)$ and the differences with respect to the complementary rule for the three scaling methods. One can see that WT-MFDFA still performs better.
  }
  \label{fig-C5}
\end{figure}
%%%%% บบบบบบบบบบบบบบบบบบบบบบบบบบบบบบบบบบบบบบบบบบบบFIG. C6
\begin{figure} [x] %[h!]
 \centering
  \includegraphics[width=11.5cm, height=13cm]  {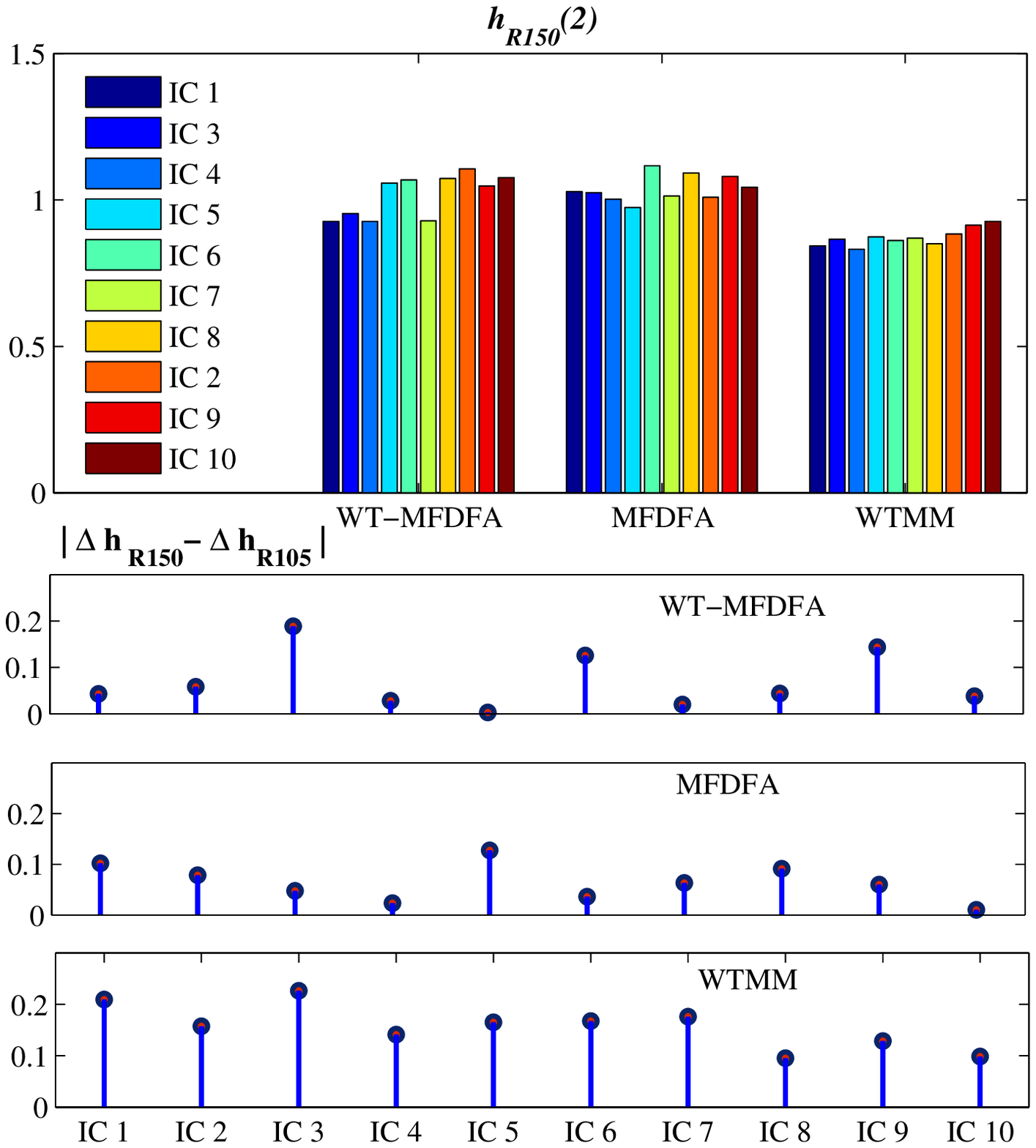}   %{13f-R90_Db2.eps}   %{R90NA_db6_db16}
   \caption{\sl \small
   Rule 150 ECA: The same as in the previous figure. WT-MFDFA is still better globally.
  }
  \label{fig-C6}
\end{figure}

\bigskip
\bigskip

%%% ============================
{\bf Appendix: Wavelet theory}\\
%%% ============================

\noindent The \emph{wavelet transform} (WT for short) of a signal $x(t)$ is given by
%..................
\begin{equation}
   W_x(a, b) = \frac{1}{\sqrt{a}} \int_{-\infty}^{\infty} x(t)
   \bar{\psi} \left(\frac{t - b}{a} \right) dt,
   \label{eq-CWT}
   \end{equation}
where $\psi$ is the analyzing wavelet, $b \in \mathbb{R}$ is a translation parameter, whereas $a \in \mathbb{R}^{+} ~ (a \neq 0)$ is a
dilation or scale parameter, and the bar symbol denotes complex conjugation.

One fundamental property that is required to analyze singular behavior is that $\psi(t)$ has enough vanishing moments~\cite{arne1, Mallat}.
   A wavelet is said to have $n$ vanishing moments if and only if it satisfies
   %%%%% ===========
   \begin{align} \label{momentos}
   \int_{-\infty}^{\infty} t^k \psi(t) dt & = 0, \quad
   \text{for ~ $k = 0, 1, \ldots , n - 1$,}\,
   %\intertext{and}
   \quad \text{and} \quad
   \int_{-\infty}^{\infty} t^n \psi(t) dt \neq 0~. %\quad \text{for ~ $k = n.$} \notag
   \end{align}
   %%%%% ===========
   %%%%% ===========
This means that a wavelet with $n$ vanishing moments is orthogonal to all polynomials up to order $n - 1$. Thus, the wavelet
transform of $x(t)$ performed by means of a wavelet $\psi(t)$ with $n$ vanishing moments is nothing but a ``smoothed version'' of the $n$--th derivative of $s(t)$ on various scales.

    Since in general the majority of data are represented by a finite number of values, we consider the orthogonal (discrete)
    wavelet case in which the wavelets are associated to orthonormal bases of $L^2(\mathbb{R})$.
    First, a discrete grid is established for the dilation and translation parameters
    such that wavelet function $\psi(t)$, and its associated scaling function $\varphi(t)$
    \cite{Mallat, Ingrid} can be expressed in the form
%% ------------------------------------
   \begin{equation}\label{eq-TO-base}
     \varphi_{m,n}(t) = 2^{m/2} \varphi(2^m t - n), \qquad
     \psi_{m,n}(t) = 2^{m/2} \psi(2^m t - n), \quad m,n\in \mathbb{Z}
   \end{equation}
   %% ------------------------------------
with $m$ and $n$ denoting the dilation and translation indices, respectively.

    Within this framework, one can write the expansion of an arbitrary signal  $x(t)$ in an orthonormal wavelet basis as follows
% =====================================
  \begin{equation}\label{eq-TO-sint}
      x(t) = \sum_n \left(a_{m_0,n} \varphi_{m_0,n}(t)+
     \sum_{m = m_0}^{M-1}d_{m,n} \psi_{m,n}(t)\right),
 \end{equation}
 % =====================================
where the scaling or approximation coefficients $a_{m,n}$ and the wavelet coefficients $d_{m,n}$ are given by
% =====================================
 \begin{equation}\label{eq-coef-a-d}
    a_{m,n} = \int x(t) \varphi_{m,n}(t) dt, \qquad
    d_{m,n} = \int x(t) \psi_{m,n}(t) dt.
 \end{equation}
 % =====================================

   In this context, a computationally efficient method to compute \eqref{eq-coef-a-d} was developed by Mallat \cite{Mallat} under the name of
multi-resolution analysis (MRA). The MRA approach provides a general method for constructing orthogonal wavelet basis and leads to the implementation of the fast wavelet transform (FWT). A multi-resolution decomposition of a signal is based on successive decomposition into a series of
approximations and details, which become increasingly coarse. The FWT calculates the scaling and wavelet coefficients at scale $m$ from the scaling coefficients at the next finer scale $m+1$ using
% =====================================
  %% ------------------------------------
     \begin{align} \label{eq-TO-pro}
     a_{m,n} & = \sum_k h[k - 2n] a_{m+1,k}, \\
       \label{eq-TO-det}
     d_{m,n} & = \sum_k g[k - 2n] a_{m+1,k},
    \end{align}
  %% ------------------------------------
where $h[n]$ and $g[n]$ are typically called low pass and high pass filters in the associated analysis filter bank.
  In fact, the signals $a_{m,n}$ and $d_{m,n}$ are the convolutions of $a_{m+1,n}$ with the filters $h[n]$ and $g[n]$ followed by a downsampling
  of factor 2 \cite{Mallat}, respectively.

  Conversely, a reconstruction of the original scaling coefficients $a_{m + 1,n}$ can be made from the following combination of the scaling and wavelet coefficients at a coarse scale
  %% ------------------------------------
  \begin{equation} \label{eq-TO-FBS}
        a_{m + 1,n} = \sum_k \left( h[2k - n] a_{m,k} + g[2k - n] d_{m,k} \right)~.
     \end{equation}
  %% ------------------------------------
  It corresponds to the synthesis filter bank. This part can be viewed as the discrete convolutions between the upsampled signal $a_{m,l}$ and the filters $h[n]$ and $g[n]$, that is, following an upsampling of factor 2 the convolutions between the upsampled signal and the filters
  $h[n]$ and $g[n]$ are calculated.

%
%%%%% &&&&&&&&&&&&&&&&&&&&&&&&&&&&&&&&&&&&&&&&&&&&&&&&&&&&&&&&&&&&

\end{document}